# Evaluation of Seed Set Selection Approaches and Active Learning Strategies in Predictive Coding

Christian J. Mahoney
e-Discovery
Cleary Gottlieb Steen & Hamilton LLP.
Washington, D.C. USA
cmahoney@cgsh.com

Nathaniel Huber-Fliflet
Data & Technology
Ankura Consulting Group, LLC
Washington, D.C. USA
nathaniel.huber-fliflet@ankura.com

Haozhen Zhao
Data & Technology
Ankura Consulting Group, LLC
Washington, D.C. USA
haozhen.zhao@ankura.com

Jianping Zhang
Data & Technology
Ankura Consulting Group, LLC
Washington, D.C. USA
jianping.zhang@ankura.com

Peter Gronvall
Data & Technology
Ankura Consulting Group, LLC
Washington, D.C. USA
peter.gronvall@ankura.com

Shi Ye
Data & Technology
Ankura Consulting Group, LLC
Washington, D.C. USA
shi.ye@ankura.com

## ABSTRACT

Active learning is a popular methodology in text classification – known in the legal domain as 'predictive coding' or 'Technology Assisted Review' or 'TAR' – due to its potential to minimize the required review effort to build effective classifiers. It is generally assumed that when building a classifier of data for legal purposes (such as production to an opposing party or identification of attorney-client privileged data), the seed set matters less as additional learning rounds are performed, thus in most existing relevant seed set studies the seed set is either built from a random document set or from synthetic documents. However, our recent empirical evaluation on a range of seed set selection strategies demonstrates that the seed set selection strategy can significantly impact predictive coding performance. It is unclear whether that conclusion applies to active learning for predictive coding. In this study, we try to answer that question by using extensive experimentation which examines the impact of popular seed set selection strategies in active learning, within a predictive coding exercise. Additionally, significant research has been devoted to achieving high levels of recall efficiently through continuous active learning strategies when there is an assumption that human review will continue until a certain recall is achieved. However, for reasons such as monetary costs, sensitivity of data (or lack thereof), or time to classify a population, this heavy human lift is often less than ideal for lawyers that are classifying a population for production to an opposing party or classifying a population for attorney-client privilege. Often the strategy is to, instead, minimize the human review effort and to classify a population efficiently with minimal human intervention. In these instances, the selection strategy may be different than what prior research suggests. In this study, we evaluate different active learning strategies against well-researched continuous active learning strategies for the purpose of determining efficient training methods for classifying large populations quickly and precisely. We study how random sampling, keyword models and clustering based seed set selection strategies combined together with top-ranked, uncertain, random, recall inspired, and hybrid active learning document selection strategies affect the performance of active learning for predictive coding. For the purpose of this study, we use the percentage of documents requiring review to reach 75% recall as the 'benchmark' metric to evaluate and compare our approaches. 75% is a commonly used recall threshold in the legal domain when using classifiers to designate documents for production. In most cases we find that seed set selection methods have a minor impact, though they do show significant impact in lower richness data sets or when choosing a top-ranked active learning selection strategy. Our results also show that active learning selection strategies implementing uncertainty, random, or 75% recall selection strategies has the potential to reach the optimum active learning round much earlier than the popular continuous active learning approach (top-ranked selection). The results of our research shed light on the impact of active learning seed set selection strategies and also the effectiveness of the selection strategies for the following learning rounds. Legal practitioners can use the results of this study to enhance the efficiency, precision, and simplicity of their predictive coding process.

## 1 Introduction

The exponential growth of electronically stored information (ESI) falling within the scope of today's large legal cases creates unique challenges for all parties involved, including clients, lawyers, and courts/tribunals/enforcement agencies. Given the volumes and complexities of ESI, litigators struggle to identify documents relevant to a case (with data populations doubling about every two years) [10], while maintaining the quality and affordability of legal document review. Companies regularly spend millions of dollars producing responsive ESI for matters in litigation, and research shows that often the majority of the costs are incurred by the review process [12]. The traditional manual review approach is often neither economically feasible nor timely enough to meet courts' or regulators' requirements. To confront these challenges, predictive coding is increasingly embraced by legal practitioners to cull through massive volumes of data for relevant information. Predictive coding, or text classification as it is referred to in the machine learning domain, uses a machine learning algorithm to train a model from a sample set, then uses the model to identify documents that are potentially relevant, which can then be isolated for legal document production or prioritized for review.

A common protocol in applying predictive coding in legal document review is to, instead of relying on a single model trained from a single seed set, train predictive coding models using an iterative approach. Following the coding of a first round of training, commonly referred to as a seed set, an initial predictive model is created – this model is used to score all the unlabeled documents. Then, a training document selection strategy is used to choose new training documents from the scored population. These documents will be reviewed, and then added to the training set to train a new version of the model. This process is repeated until the goal of manually finding enough relevant documents during an active learning review is met (a strategy called Continuous Active Learning, or "CAL") or until the performance of the latest model meets an acceptable recall threshold with an acceptable amount of precision. Once this level is met the document-level scoring from the classifier is used to make a relevance determination on the remaining unreviewed documents in the population (a strategy called Simple Active Learning, or "SAL"). Existing studies show that active learning approaches provide an advantage by finding as many relevant documents as possible while spending minimal review efforts [3]. However, these studies assume the human review of all documents identified as relevant by the predictive model and focus on how best to expedite this process through continuous prioritization of relevant documents until target recall thresholds are achieved [4]. This group of collaborators finds that though there are real world legal matters where such human review is excessively costly or time consuming, there are a lack of studies that focus on SAL and how to most efficiently train an active learning model that efficiently achieves a high level of recall with minimal human review of training documents.

In certain situations, particularly where minimizing either the time or cost to classify a data set is paramount, this can be a more desirable approach than a Continuous Active Learning protocol that reprioritizes documents round after round until a desired recall is achieved through human review. There are two critical aspects of this kind of protocol. One aspect concerns the initial seed set used to train the first-round model – whether seed sets selected using different approaches ultimately have a significant impact on an active learning model. The other aspect concerns the impact of how additional training documents are chosen and added to improve model performance. A more thorough understanding of these two aspects will provide guidance to legal practitioners to make decisions in managing the predictive coding process that will help to minimize the amount of time and cost to develop a highly effective model.

In this paper, we report our empirical studies on the impact of seed set selection and active learning document selection strategies on predictive coding for legal document review. We use four fully coded or labeled data sets prepared in response to production requests in actual legal matters spanning across different industries. For each of these data sets, we utilize their keyword search terms for testing seeding and iterative training strategies. We conducted roughly 115,000 rounds of predictive coding experiments to study how different seed set selection and active learning document selection strategies affect the performance of predictive coding. Our paper is organized as follows. (i) We first review existing research related to seed set selection and active learning document selection strategies. (ii) We then lay out our methodology, including the seed set selection and active learning document selection strategies, as well as our research questions. (iii) Next, we introduce the data sets used in our experiments, our experimental procedure, and our evaluation metrics. (iv) Finally, we discuss our experimental results and conclude the paper with key insights from our study and describe future work.

## 2 Related Work

The seed set, as the initial training set for predictive coding, has created significant debates in the legal domain. One of these debates centers around how the seed set, or initial training set of documents, should be generated. In our research, we focus on the best strategies to generate seed sets.

There is no established consensus on seed set selection sampling methods. Two major seed set selection methods are: random sampling and judgmental sampling. Schieneman et. al. argue that the seed set should be "representative of the collection" thus based on random sampling such that the predictive coding process would result in adequate recall and that judgmental sampling could potentially "bias" results [14]. In contrast, Cormack et al. [4] propose the use of a synthetic seed document, e.g. constructed from topic descriptions, in their AutoTAR protocol. Pickens et al. [13] studied manual seeding in the TREC Total Recall Track and found that initial seeding conditions had impact on task outcomes. In our previous work [11], we studied the effect of different seed set selection strategies in predictive coding, and empirically demonstrated that complex seed set selection techniques with the purpose of ensuring the diversity of the seed set or increasing its richness only provides modest improvement when compared to the random sampling.

In active learning protocols, a key component is the method of selecting additional training documents after each round. The seminal work by Lewis et al. [8, 9] showed that choosing additional training documents closest to a score of .5 (on a scale of 0 to 1), that are the area that Lewis describes as most uncertain to the classifier, produces an effective classifier quicker than other selection strategies. In their original paper on the Continuous Active Learning protocol, Cormack et al. [3] compared three active learning document selection strategies: (i) select top scored documents (most commonly associated with CAL); (ii) select documents of which the learning algorithm is most uncertain in making a relevance call (most commonly associated with SAL); and (iii) select documents randomly (most commonly associated with Simple Passive Learning, or SPL). Their paper demonstrated that the CAL training selection strategy consistently outperformed other approaches in finding the most relevant documents with minimal review efforts. Chhatwal et al. [2] also studied the same three active learning document selection strategies, Top-Ranked, Uncertain, and Random applied to real legal matter data sets. This study revealed that always selecting the highest-scoring documents as additional training documents may not be the most efficient approach because round by round the model's performance may not improve. Both conclusions are understandable if we appreciate the dual purpose inherent in active learning: (i) quickly find as many relevant documents as possible; (ii) train an effective final model using as few rounds as possible. The conflicting conclusions of the two studies are due to evaluating the selection strategies differently. In Cormack's work, the performance was evaluated using only the training set, namely the documents that were selected. In our work, the performance was evaluated on both the documents selected and the documents classified by the model. Recently, there are new efforts in experimenting with retraining strategies in CAL. Ghelani et al. [5] compared retraining with exponentially increased or static top-scored documents, as well as partial retraining, precision-based, and recency weighted retraining strategies, and show that CAL can achieve higher recall when retraining more frequently.

## 3 Training Document Selection

In this section, we introduce both seed set document selection and active learning document selection strategies.

### 3.1 Seed Set Selection Methods

In our previous paper [11], we studied the predictive coding performance of the following seed set selection strategies.
- Random Sampling (*random*): generate a random sample of documents from the corpus of all documents.
- Stratified Keyword Sampling (*keyword_method1*): select an equal number of documents from the document hits of each keyword developed by counsel for the purpose of identifying responsive information.
- Weighted Stratified Keyword Sampling (*keyword_method2*): select a number of documents from document hits of each keyword proportional to the hits size.
- Clustering Sampling (*cluster_method1*): select an equal number of documents from each cluster. We use a variant of the K-Means clustering algorithm to create a cluster set of three branches to a depth of five layers for each data set.
- Weighted Clustering Sampling (*cluster_method2*): select a number of documents from each cluster proportional to the cluster size.

More detailed description of these seed set selection methods can be found in [11].

### 3.2 Active Learning Selection Strategies

Six active learning selection strategies were studied in this research.
- Top-Ranked (*TOP*): select documents with the highest scores assigned by the model.
- Uncertain (*MID-50*): select documents nearest to the score of 0.5 (in either direction from .5), which is the score indicating highest uncertainty prescribed by our model.
- MID at 75% recall (*MID_75RC*): select documents nearest the cut-off score (in either direction from the cut off score) resulting in a recall of 75% of all responsive documents.
- Random (*RAND*): select documents randomly from all the documents scored by the model.
- 80% Top scored + 20% random (*80TOP20RD*): select 80% of the documents with the highest scores assigned by the model and 20% of the documents randomly from the rest.
- 20% Top scored + 80% random (*20TOP80RD*): select 20% of the documents with the highest scores assigned by the model and 80% of the documents randomly from the rest.

It should be noted the MID_75RC strategy is a novel strategy that we have not seen in any literature. The reason we used 75% recall is that in real-world legal document reviews, a recall of 75% is a commonly used minimum performance metric. In practice, this strategy can be implemented as selecting documents with scores nearest to the cut-off score for 75% recall derived from a statistically representative sample set – essentially implementing an initial validation set (or control set) is required to implement this strategy in a real-world scenario. As an example, a control set of 2,000 documents is isolated and coded by human reviewers and has a richness of 20%, resulting in 400 relevant documents within the random sample. The classifier would achieve an estimated 75% recall by identifying the cutoff score at which 300 of the 400 relevant documents are identified by the classifier. For purposes of this study, we have used fully coded document populations in order to eliminate the uncertainty involved with this type of recall estimate.

Our research empirically compared different seed set selection strategies combined with different active learning document selection strategies. Specifically, we address the following questions:
1. What effect do different seed set selection strategies have on the active learning process?
2. What effect do different active learning document selection strategies have on the predictive coding process?
3. How do seed set selection strategies impact the effectiveness of active learning selection strategies?
4. Are there combinations of seed set and active learning strategies that consistently outperform other strategies when an emphasis is placed on objectives most commonly associated with a SAL approach (namely minimizing the amount of human review, time, and costs in isolating a precise population that achieves a certain recall threshold)?

## 4 Experiments

In this section, we first introduce the data sets we used in the empirical study, and then we discuss the experimental procedure and evaluation metrics. We report the experimental results in the next section.

### 4.1 Data Sets

We conducted experiments on four data sets from confidential, non-public, real legal matters across various industries such as social media, communications, construction, and security. We chose matters with data sets that ranged from around 300,000 to 500,000 documents in order to execute our experiments within a reasonable time period. The richness, or positive class rate, of the four data sets ranged from approximately 4% to 39%. Attorneys reviewed all documents in the four data sets over the course of the legal matter and their coding (labels) provided the ability to fully evaluate the performance of the models. Tables 1A, 1B, 1C, and 1D provide the details for the four data sets, respectively. The details include descriptions, sizes, attorney coding statistics, and statistics about keyword terms on the data sets. The predictive coding objective for Data Sets A, B, and C was to identify privileged communications between attorneys and clients. The objective for Data Set D was to identify documents responsive to production requests from the opposing party in the matter.

The recall of the keyword hits is around 93% for the privileged data sets and 34% for the responsive data set. As comparing keywords-based and predictive coding approaches for legal document review is beyond the scope of this paper, readers that are interested in this subject can read our previous related papers [6, 7].

Table 1A: Privilege Data Set Statistics

| Data Sets | Total Documents | Privileged Documents | Not Privileged Documents | Richness |
|---|---|---|---|---|
| Project A | 308,621 | 46,730 | 261,891 | 15.14% |
| Project B | 393,745 | 14,307 | 379,438 | 3.63% |
| Project C | 277,412 | 38,834 | 238,578 | 14.00% |

Table 1B: Responsive Data Set Statistics

| Data Sets | Total Documents | Responsive Documents | Not Responsive Documents | Richness |
|---|---|---|---|---|
| Project D | 412,880 | 159,304 | 253,576 | 38.58% |

Table 1C: Privilege Keyword Statistics

| Data Sets | Total Documents | Keywords | Documents Hit by Keywords | Privileged Documents Hit by Keywords | Keyword Hit Percentage |
|---|---|---|---|---|---|
| Project A | 308,621 | 808 | 193,017 | 43,847 | 62.54% |
| Project B | 393,745 | 4,211 | 368,506 | 13,571 | 93.59% |
| Project C | 277,745 | 509 | 159,900 | 36,234 | 57.57% |

Table 1D: Responsive Keyword Statistics

| Data Sets | Total Documents | Keywords | Documents Hit by Keywords | Responsive Documents Hit by Keywords | Keyword Hit Percentage |
|---|---|---|---|---|---|
| Project D | 412,880 | 23 | 81,362 | 53,611 | 19.71% |

### 4.2 Experiment Procedure

We conducted an empirical study on the effect that seed set and active learning document selection strategies have on the performance of a predictive coding process.

The same set of experiments were performed on each of the four data sets. For each data set, all of the five seed set selection strategies and all of the six active learning document selection strategies were tested. In total there were 30 combinations of seed set selection and active learning document selection strategies for each data set. In all experiments, the seed set included 500 training documents, and an additional 250 training documents are selected in each round of active learning. Table II shows the richness of the seed sets for the four data sets. From the table, we can see that the Random seed set selection method generally has similar richness as that of the overall data set, while seed sets derived from keyword search have higher richness than the overall data set.

Table 2: Richness of seed sets (%)

| Data Sets | Random | Keyword Method 1 | Keyword Method 2 | Cluster Method 1 | Cluster Method 1 |
|---|---|---|---|---|---|
| Project A | 14.8 | 40.2 | 43.4 | 15.0 | 15.2 |
| Project B | 3.6 | 6.6 | 6.8 | 3.8 | 3.2 |
| Project C | 11.8 | 36.2 | 34.4 | 14.2 | 12.8 |
| Project D | 40.4 | 70.2 | 73.8 | 40.2 | 38.6 |

Our experimental procedure was:

1. First, use the selected seed set sampling method to determine an initial training set of 500 documents.
2. Train a model with the selected seed set using the same underlying machine learning algorithm (logistic regression) and text processing parameters.
3. Then, score the entire data set, excluding any document used in training.
4. Next, select an additional 250 documents using one of the active learning document selection strategies.
5. Finally, add these new training documents, train a new model, and repeat steps 3, 4, and 5 until there are no more documents left to be scored or the minimum performance of the model is achieved.

We used Logistic Regression as the machine learning algorithm due to its consistent high performance across different settings over various data sets demonstrated in previous studies [1, 2]. Other text processing parameters we used for modeling were, bag of words with 1-gram, normalized frequency, and 20,000 tokens were used as features.

In each round of our experiments, the entire data set was used either in training or scoring, which means on average more than 300,000 documents were used in training or scoring. The total number of models trained in our experiments was: 114,933. We leveraged the Apache Lucene search engine library to build full text indices of the data sets to speed up the training and scoring processes.

### 4.3 Evaluation Metrics

Our performance metric measured the percentage of documents requiring review to achieve the targeted recall level. In the common passive learning scenario, this metric can be calculated on a validation set and does not consider documents that are reviewed for training because these documents typically have a negligible impact when attempting to achieve the desired recall performance. In an active learning scenario, as rounds increase, the number of documents reviewed for training and used to develop the model could constitute a considerable portion of the population requiring review. Therefore, performance metrics in our experiments were computed after each active learning round using two sets of documents. The (i) first set contained the documents that were selected and reviewed during training. The (ii) second set contained the documents categorized as Responsive or Privileged by the predictive model after each round, namely the documents with probability scores greater than or equal to the predictive model's cut-off score. The documents with scores greater than or equal to the cut-off score are the documents that attorneys would consider producing to an opposing party, for assertions of privilege, or in some instances for review because they are likely responsive, or in the case of privilege, may contain content that would allow for the assertion of claims of privilege.

We can use an example to illustrate the calculation of these measures. Project A has 308,621 documents, of which 46,730 are positive. Using the random seed selection strategy, we would select a seed set with 74 positive documents and 426 negative documents. Now suppose we use the TOP active learning document selection strategy, which selects 250 documents with the highest scores to add to the training set. Lastly, assume that after ten rounds the training set contains 2,491 positive documents and 509 negative documents, in total 3,000 documents. Examining the document scores after ten rounds in this example, we find that if we choose 24.7 as the cut-off score, there are 32,558 positive documents above this cut-off score. 32,558 + 2,491 = to 35,049, represents 75% of all the responsive documents (46,730) in this data set. The total population of documents requiring review is then established by adding all the documents with scores above 24.7 (82,206) to all the training documents (3,000) divided by the total population size (308,621). This would equal: 27.6%.

## 5 Results and Discussion

The total number of our experimental parameter combinations was 120. These parameters include: data set, seed set selection method, and the active learning document selection strategy. On average there were roughly 1,000 rounds of experiments generated for each combination. To save space in this paper, we only present the most interesting results.

### 5.1 The Impact of Seed Set Selection Approaches

Figures 1 displays the percentage of documents requiring review to achieve 75% recall for different seed set selection strategies. Active learning strategies were fixed to TOP and MID_75RC on Projects C and D and the first 100 rounds of experiments are shown. Figure 2 details the percentage of documents requiring review to achieve 75% recall for different seed set selection strategies. The RAND active learning document selection strategy was fixed on Project B and C and the first 100 rounds of experiments are shown. In general, these figures show that seed set selection strategies have very modest impact on the performance of the active learning strategies, especially after many rounds of active learning. These results were expected when using seed sets with a small number of documents (e.g., 500) because the initial impact of the seed set selection strategy likely degrades over training rounds. Therefore, it may be worthwhile to experiment with seed sets of larger document sizes in the future. However, from these results, we do find two salient aspects about the impact of the seed set selection approach. First, among the different active learning strategies, the TOP strategy is the most sensitive to the seed selection strategy; we can see more apparent performance difference across the different seed set selection strategies (Figure 1). This implies that in the very popular Continuous Active Learning protocol, the seed set selection strategy has an impactful role and should be considered carefully. Second, curves in Project B – a matter with 3.6% richness – show that the seed set selection strategy had a greater impact on a low richness population and that judgmental seed set selection strategies using keywords or clustering outperform randomly selected seed set documents in the early rounds (Figure 2).

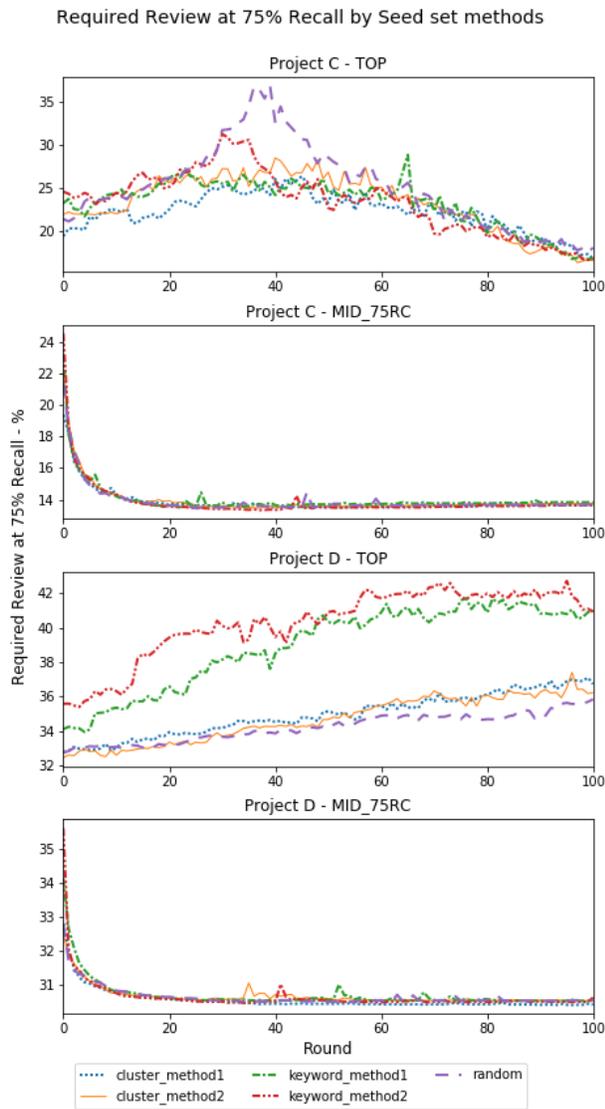

**Figure 1: Required Review at 75% Recall for the five Seed Set Methods with TOP, MID_75RC Active Learning Strategies on Project C and D (First 100 Rounds)**

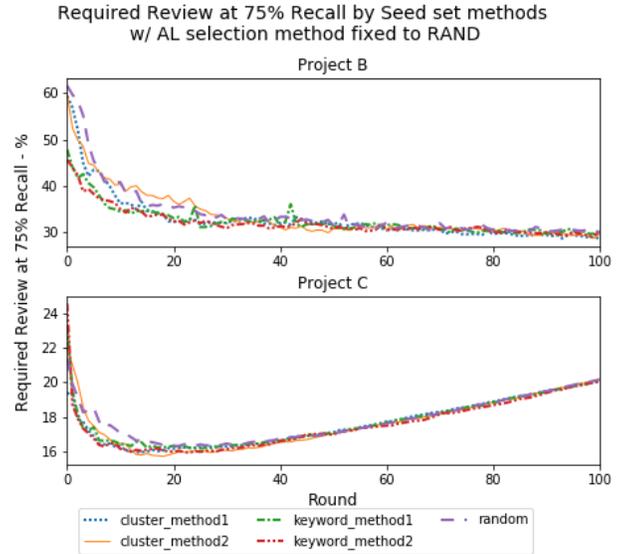

**Figure 2: Required Review at 75% Recall for the five Seed Set Methods with RAND Active Learning Strategy on Project B and C (First 100 Rounds)**

## 5.2 The Impact of Active Learning Strategies

Figure 3 shows the performance differences among TOP, MID-50, MID_75RC, and RAND with the seed set selection method fixed to *random*, over learning rounds of experiments until the optimum round is reached. These results confirm the findings in our previous research [1], i.e. active learning selection strategies such as uncertain sampling (MID-50) and random selection (RAND) can generate an effective model within fewer rounds than the popular TOP strategy. Moreover, we find that the MID_75RC strategy, a novel active learning strategy proposed for the first time in this paper, performs the best in almost all the scenarios. This indicates that selecting documents nearest to the cut-off score for 75 percent recall would be the most effective active learning strategy, when attempting to achieve 75 percent recall.

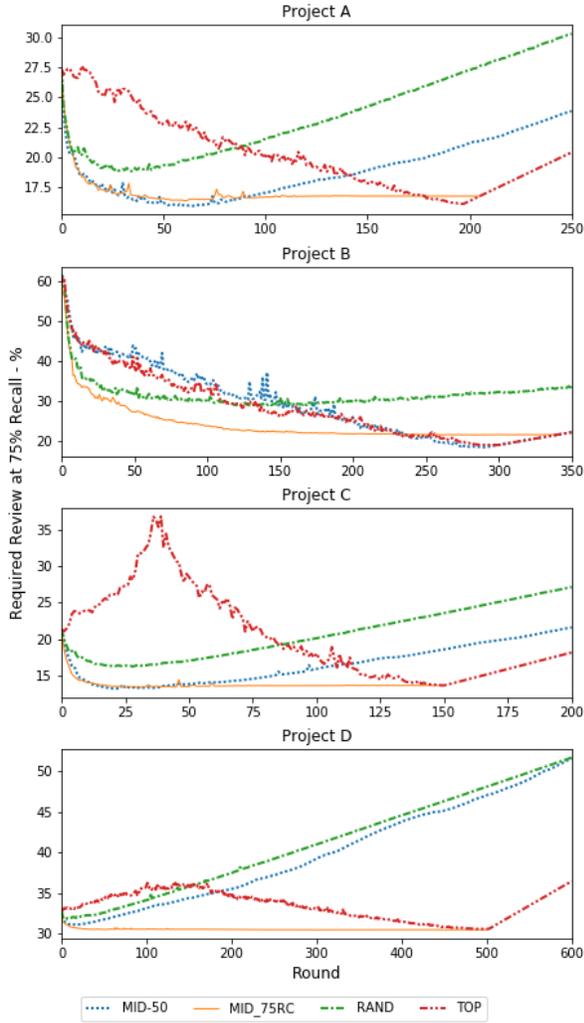

**Figure 3:** Required Review at 75% Recall TOP, MID-50, MID_75RC and RAND Active Learning Strategies with *random* Seed Set Selection Method

The performance difference between the TOP strategy and the MID_75RC strategy is even more clear when we look closely into the plots of the first 100 rounds. Table 3 shows in the first 50 rounds and the MID_75RC strategy consistently requires less review than the TOP strategy across all projects. The maximum saving would be close to 20 percent in Project C. In practice, this has a significant impact on the predictive coding process and should be considered by legal teams to help reduce review costs.

Table 3: Required Review at 75% Recall for TOP and MID_75RC Active Learning Strategies (First 50 Rounds Every 10 Rounds)

| Data Set | Round | TOP | MID_75RC | Difference |
|---|---|---|---|---|
| Project A | 10 | 28% | 18% | 9% |
| | 20 | 25% | 17% | 8% |
| | 30 | 26% | 17% | 9% |
| | 40 | 24% | 17% | 7% |
| | 50 | 23% | 17% | 6% |
| Project B | 10 | 47% | 35% | 12% |
| | 20 | 44% | 32% | 11% |
| | 30 | 43% | 31% | 12% |
| | 40 | 40% | 29% | 11% |
| | 50 | 39% | 28% | 11% |
| Project C | 10 | 24% | 14% | 10% |
| | 20 | 26% | 14% | 13% |
| | 30 | 32% | 14% | 18% |
| | 40 | 33% | 14% | 19% |
| | 50 | 29% | 14% | 15% |
| Project D | 10 | 33% | 31% | 2% |
| | 20 | 33% | 31% | 3% |
| | 30 | 34% | 31% | 3% |
| | 40 | 34% | 30% | 3% |
| | 50 | 34% | 30% | 4% |

### 5.3 Optimum Performance Round Analysis

We define the optimum performance round as the round in which the amount of review required to reach 75 percent recall is the earliest. After some analysis, we found the dominant factor in reaching the optimum performance round is the active learning strategy and not the seed set selection strategy. In Table 4A through 4D, we compiled the optimum performance round of each active learning strategy for the four data sets. We can see that strategies such as RAND, MID-50 or MID_75RC consistently take fewer rounds to reach the optimum performance round. Moreover, if a satisficing goal is set to a review percentage within 5%, 10% or 15% of the optimum performance, we can see that those strategies require fewer rounds to reach the goal.

Table 4A: Project A Optimum Performance Rounds

| Active Learning Strategy | Review Percentage | Optimum Round | 1st Round within 5% of Op. Perf. | 1st Round within 10% of Op. Perf. | 1st Round within 15% of Op. Perf. |
|---|---|---|---|---|---|
| TOP | 15.90 | 192 | 167 | 145 | 113 |
| MID-50 | 15.71 | 60 | 33 | 21 | 13 |
| MID_75RC | 16.24 | 74 | 19 | 12 | 8 |
| RAND | 18.77 | 21 | 4 | 2 | 2 |
| 80TOP20RD | 18.36 | 213 | 80 | 27 | 10 |
| 20TOP80RD | 19.09 | 50 | 7 | 3 | 2 |

Table 4B: Project B Optimum Performance Rounds

| Active Learning Strategy | Review Percentage | Optimum Round | 1st Round within 5% of Op. Perf. | 1st Round within 10% of Op. Perf. | 1st Round within 15% of Op. Perf. |
|---|---|---|---|---|---|
| TOP | 18.58 | 279 | 263 | 235 | 206 |
| MID-50 | 18.59 | 290 | 258 | 236 | 218 |
| MID_75RC | 20.88 | 263 | 168 | 117 | 92 |
| RAND | 27.79 | 123 | 85 | 41 | 22 |
| 80TOP20RD | 21.20 | 321 | 272 | 225 | 194 |
| 20TOP80RD | 27.50 | 181 | 106 | 70 | 53 |

Table 4C: Project C Optimum Performance Rounds

| Active Learning Strategy | Review Percentage | Optimum Round | 1st Round within 5% of Op. Perf. | 1st Round within 10% of Op. Perf. | 1st Round within 15% of Op. Perf. |
|---|---|---|---|---|---|
| TOP | 13.56 | 148 | 130 | 118 | 107 |
| MID-50 | 13.22 | 25 | 11 | 7 | 5 |
| MID_75RC | 13.33 | 37 | 12 | 6 | 4 |
| RAND | 15.73 | 18 | 6 | 4 | 2 |
| 80TOP20RD | 15.30 | 156 | 108 | 50 | 21 |
| 20TOP80RD | 16.05 | 29 | 6 | 3 | 2 |

Table 4D: Project D Optimum Performance Rounds

| Active Learning Strategy | Review Percentage | Optimum Round | 1st Round within 5% of Op. Perf. | 1st Round within 10% of Op. Perf. | 1st Round within 15% of Op. Perf. |
|---|---|---|---|---|---|
| TOP | 30.47 | 494 | 350 | 0 | 0 |
| MID-50 | 31.02 | 13 | 0 | 0 | 0 |
| MID_75RC | 30.36 | 332 | 1 | 0 | 0 |
| RAND | 31.65 | 8 | 0 | 0 | 0 |
| 80TOP20RD | 31.90 | 5 | 0 | 0 | 0 |
| 20TOP80RD | 31.58 | 6 | 0 | 0 | 0 |

\* Round 0 means the initial round.

## 6 Conclusions

Our experiment results show that seed set selection strategies have little impact on the active learning process. However, for low richness projects, keyword-based seed set selection strategies have more apparent effect. Also, the popular TOP active learning strategy is the most sensitive strategy to different seed selection methodologies.

Our results also show that choosing documents nearest to the cut-off score determined by reaching a 75 percent document recall potentially result in a high performing model quickly. When excluding data sets with extremely low richness (such as Project B), this training methodology results in significantly higher performing models in early training rounds, such as round 10 or round 20, rounds that are often associated with stopping points for Simple Active Learning models. In fact, in all three of our data sets that had richness above 10 percent, using the MID_75RC active learning strategy resulted in achieving performance within roughly 10 percent of the optimum model performance within 10 rounds of active learning. In theory, focusing training around the dynamic cut-off score from round to round makes sense. Documents just above the cut-off score should be the documents included as positives by the model with the least amount of certainty, so there should be the most opportunity to improve precision by improving performance by classifying the features within these documents. Documents just below the cut-off score should be the documents excluded as negatives by the model that have the highest amount of richness in the excluded population, so there should be the most opportunity to improve recall by classifying the features within these documents. It will be interesting to continue to test these assumptions and study this strategy both in data sets with low richness and in utilizing other cut-off scores to meet different recall objectives or thresholds, such as those prescribing 50 percent or 90 percent recall. It should be noted that in our current study we fixed the seed set size at 500 and the additional number of training documents in each round at 250. In future studies, we intend to examine seed sets of larger sizes or various sizes of additional active learning training documents.

The results provide practical techniques that legal practitioners can use to enhance their active learning predictive coding processes, as well as influencing their training document selection strategies for passive learning approaches.